\documentclass{appolb} 
\usepackage{lineno}
\usepackage{xspace}
\usepackage{comment}
\usepackage{subfigure}
\usepackage{graphicx} 
\usepackage{amsmath} 
\usepackage{amssymb} 
\usepackage{hyperref} 
\usepackage{cite} 

\newcommand{\jpsi} {\ensuremath{{\mathrm J}/\psi}\xspace}
\newcommand{\psip} {\ensuremath{\psi'}\xspace}
\newcommand{\rhozero} {\ensuremath{\rho^0}\xspace}

\newcommand{\PbPb}         {\mbox{Pb--Pb}\xspace}

\newcommand{\pPb}          {\mbox{p--Pb}\xspace}

\newcommand{\nineH}        {$\sqrt{s}~=~0.9$~Te\kern-.1emV\xspace}
\newcommand{\seven}        {$\sqrt{s}~=~7$~Te\kern-.1emV\xspace}
\newcommand{\eight}        {$\sqrt{s}~=~8$~Te\kern-.1emV\xspace}
\newcommand{\twoH}         {$\sqrt{s}~=~0.2$~Te\kern-.1emV\xspace}
\newcommand{\twosevensix}  {$\sqrt{s}~=~2.76$~Te\kern-.1emV\xspace}
\newcommand{\five}         {$\sqrt{s}~=~5.02$~Te\kern-.1emV\xspace}
\newcommand{\fiveExactly}  {$\sqrt{s}~=~5$~Te\kern-.1emV\xspace}
\newcommand{\twosevensixnn}{$\sqrt{s_{\mathrm{NN}}}~=~2.76$~Te\kern-.1emV\xspace}
\newcommand{\fivenn}       {$\sqrt{s_{\mathrm{NN}}}~=~5.02$~Te\kern-.1emV\xspace}

\newcommand{\GeVc}         {Ge\kern-.1emV/$c$\xspace}
\newcommand{\MeVc}         {Me\kern-.1emV/$c$\xspace}
\newcommand{\TeV}          {Te\kern-.1emV\xspace}
\newcommand{\GeV}          {Ge\kern-.1emV\xspace}
\newcommand{\GeVtwo}       {Ge\kern-.1emV$^2$\xspace}
\newcommand{\MeV}          {Me\kern-.1emV\xspace}
\newcommand{\GeVmass}      {Ge\kern-.2emV/$c^2$\xspace}
\newcommand{\MeVmass}      {Me\kern-.2emV/$c^2$\xspace}

\begin{document}

\title{Overview of the latest ALICE UPC and photonuclear results%
  \thanks{Presented at ``Diffraction and Low-$x$ 2024'', Trabia (Palermo, Italy), September 8-14, 2024.}}

\author{Simone Ragoni, for the ALICE collaboration%
  \address{Creighton University,\\ 2500 California Plz, Omaha,\\ NE 68178, United States, USA\\ \texttt{simone.ragoni@cern.ch}}
}

\date{\today}

\maketitle

\begin{abstract}
Ultra-peripheral collisions (UPC) are events characterised by large impact parameters between the two projectiles, larger than the sum of their radii. In UPCs, the protons and ions accelerated by the LHC do not interact via the strong interaction and can be regarded as sources of quasireal photons. Using the Run 2 data, the ALICE Collaboration has carried out various measurements of different final-state systems, such as exclusive four pion photoproduction as well as photoproduction of $K^+K^-$ pairs, measured for the first time in ultra-peripheral collisions. In addition, vector meson production in \PbPb provides the unique opportunity to carry out an analogy of the double-slit experiment at femtometre scales, owing to the interference between the production sources of the two lead nuclei. These results and prospects for UPC measurements using Run 3 data will be presented.
\end{abstract}

\section{Introduction}
Vector meson photoproduction in \pPb and \PbPb ultra-peripheral collisions (UPCs) \cite{Baltz:2007kq} is actively being studied at the CERN LHC. In these events, a photon from one of the two nuclei interacts with a colourless object from the other nucleus, resulting in the production of a vector meson. The ALICE Collaboration has studied \jpsi \cite{ALICE:2023jgu, ALICE:2023mfc}, \psip, \rhozero \cite{ALICE:2020ugp}, $K^+K^-$ \cite{ALICE:2023kgv} and more light vector-meson photoproduction. The interest for these processes is growing since they shed light on nuclear shadowing, gluon saturation, and more recently  gluonic hotspots, and their dependence on energy \cite{ALICE:2023gcs, Cepila:2023dxn}. 
\section{ALICE results using Run 2 (2015–2018) data}

\begin{figure}[hb]
\begin{center}
\includegraphics[width=0.56\columnwidth]{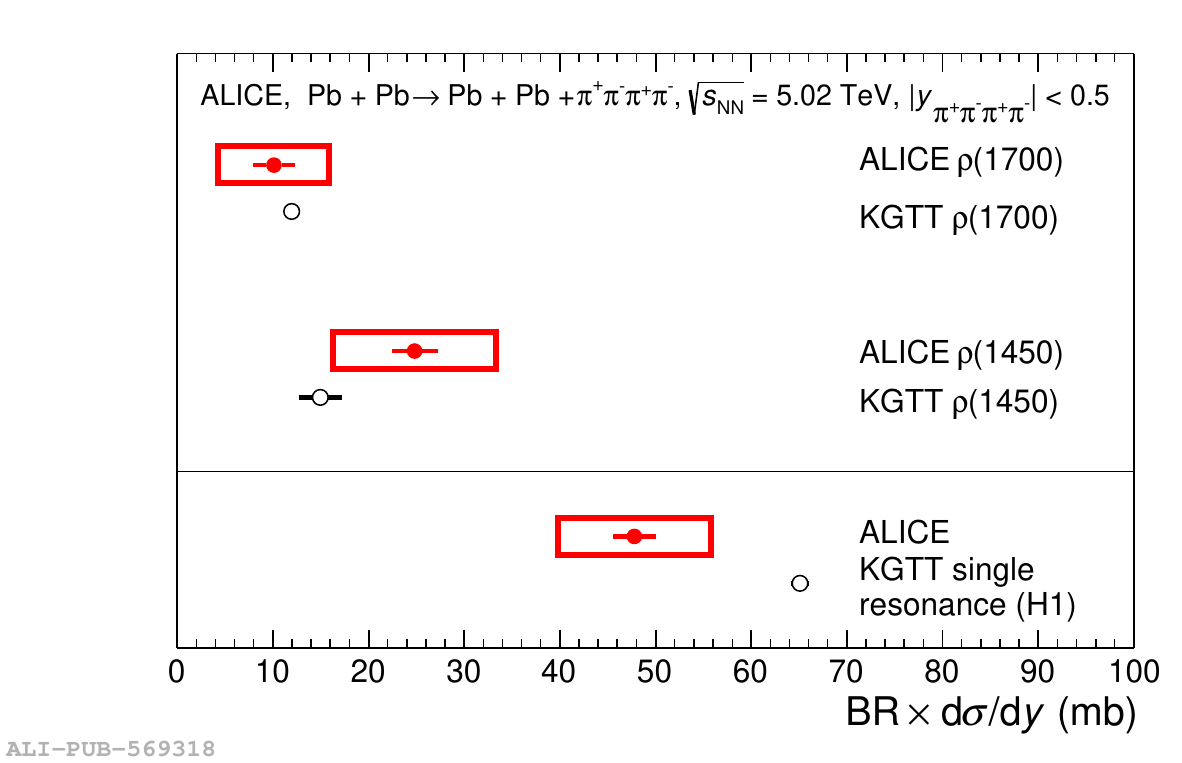}
\end{center}
\caption{\label{fig:4pi-1} Comparison of the $\rho(1450)$ and $\rho(1700)$  cross sections (upper panel), and single-resonance cross section (lower panel),  as extracted from the fits to the invariant-mass distribution of exclusive $4\pi$ production \cite{ALICE:2024kjy} and comparison with the theoretical predictions for a one and two-resonance model \cite{Klusek-Gawenda:2020gwa}.}
\end{figure}

\begin{figure}[hb]
\begin{center}
\includegraphics[width=0.6\columnwidth]{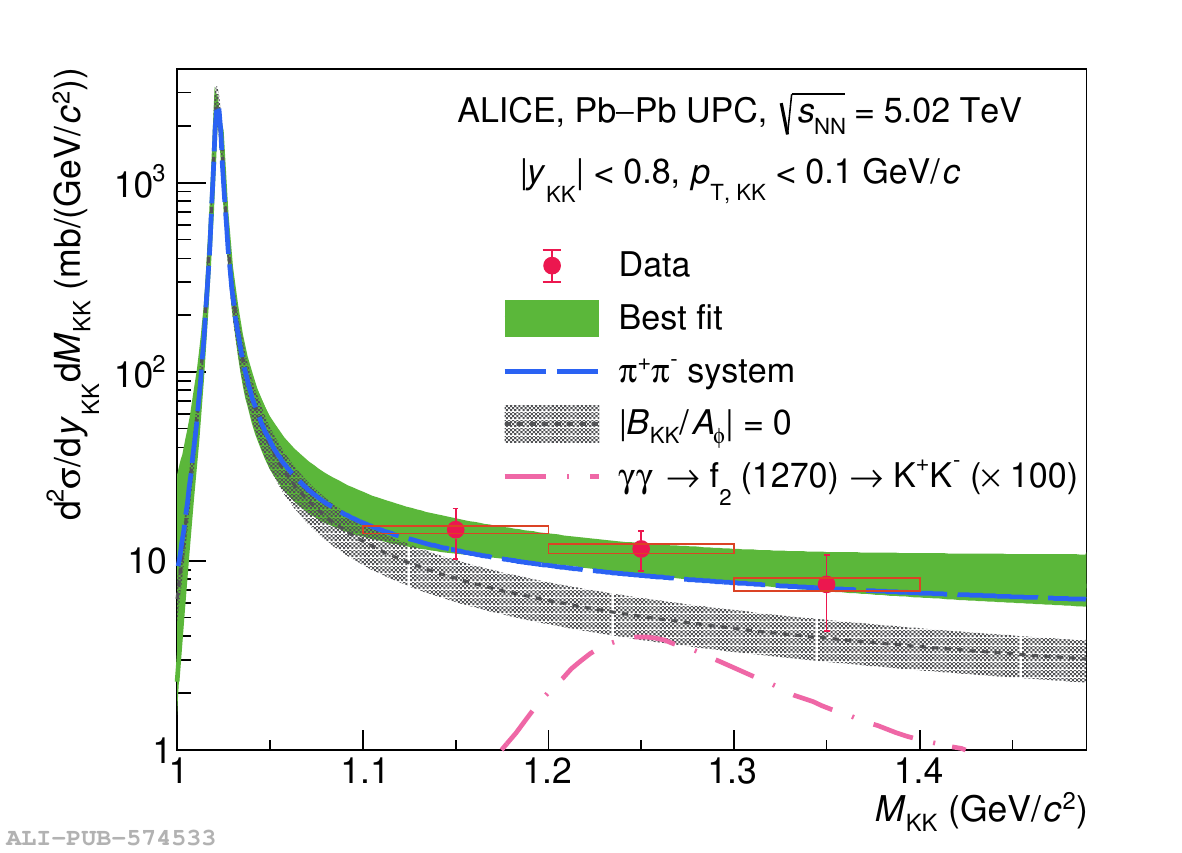}
\end{center}
\caption{\label{fig:kk} Cross sections for $K^+K^-$ photoproduction as a function of the $K^+K^-$ invariant mass as measured by ALICE \cite{ALICE:2023kgv}. The data are compatible with a cocktail of resonant and non-resonant contributions.}
\end{figure}

ALICE has recently presented results for exclusive charged $4\pi$ photoproduction \cite{ALICE:2024kjy}. The $4\pi$ invariant mass distribution is fitted using either a single Breit-Wigner or a combination of contributions representing excited \rhozero states, i.e. $\rho(1450)$ and $\rho(1700)$. The cross sections are compared to theoretical models \cite{Klusek-Gawenda:2020gwa} and are shown in Fig.~\ref{fig:4pi-1}. The computation for the combination of $\rho(1450)$ and $\rho(1700)$ (upper panel) is in better agreement with the data.
The exclusive $K^+K^-$ photoproduction results \cite{ALICE:2023kgv} show that the sample is a cocktail of resonant and non-resonant contributions, as shown in Fig.~\ref{fig:kk}. The $K^+K^-$ invariant mass distribution is measured for states above 1.1 \GeVmass, away from the $\phi(1020)$ peak, clear sign that the energy loss in the tracking material is too significant for the decay kaons from the $\phi(1020)$ to be able to reach the Time Projection Chamber (TPC) using the Inner Tracking System (ITS)  that was installed in ALICE during Run 1 and 2. A new ITS was installed during the Long Shutdown 2 \cite{Reidt:2021tvq}, with reduced material budget and higher precision.

\begin{figure}[hb!]
\begin{center}
\includegraphics[width=0.6\columnwidth]{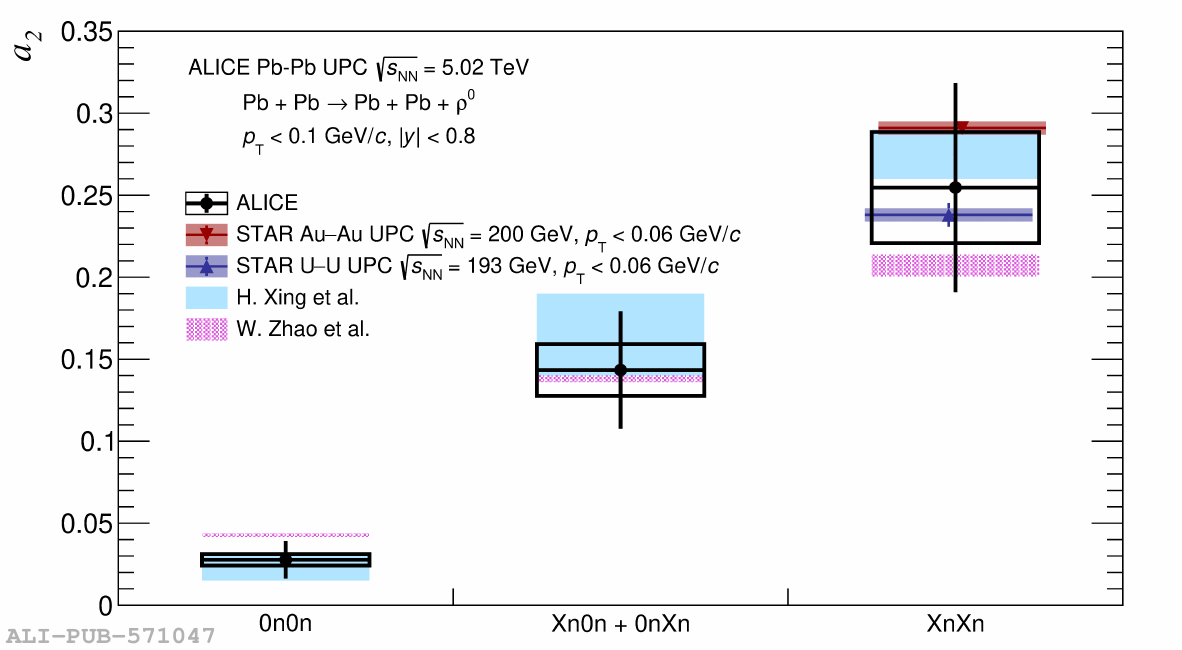}
\end{center}
\caption{\label{fig:azimuth} Impact-parameter dependent azimuthal anisotropy measured in coherent \rhozero photoproduction \cite{ALICE:2024ife}. The amplitude $a_2$ of the modulation increases as the impact parameter becomes smaller.}
\end{figure}

The ALICE Collaboration has also recently provided new results concerning the impact-parameter dependent azimuthal anisotropy in UPCs, which was measured in coherent \rhozero production \cite{ALICE:2024ife}. The results are shown in Fig.~\ref{fig:azimuth}. The amplitude $a_2$ of the modulation  increases as the impact parameter becomes smaller, which is  achieved in UPCs by separating the data set in neutron emission classes \cite{Guzey:2013jaa}. From the 0n0n to the XnXn class, the impact parameters lower from a median of about 49 fm to about 18 fm \cite{Baltz:2002pp, ALICE:2024ife}.

\section{ALICE future opportunities using Run 3 and 4 data}
\begin{figure}[ht!]
	\begin{center}
		\subfigure[]{
			\label{fig:rho-run2}
			\includegraphics[width=0.46\textwidth]{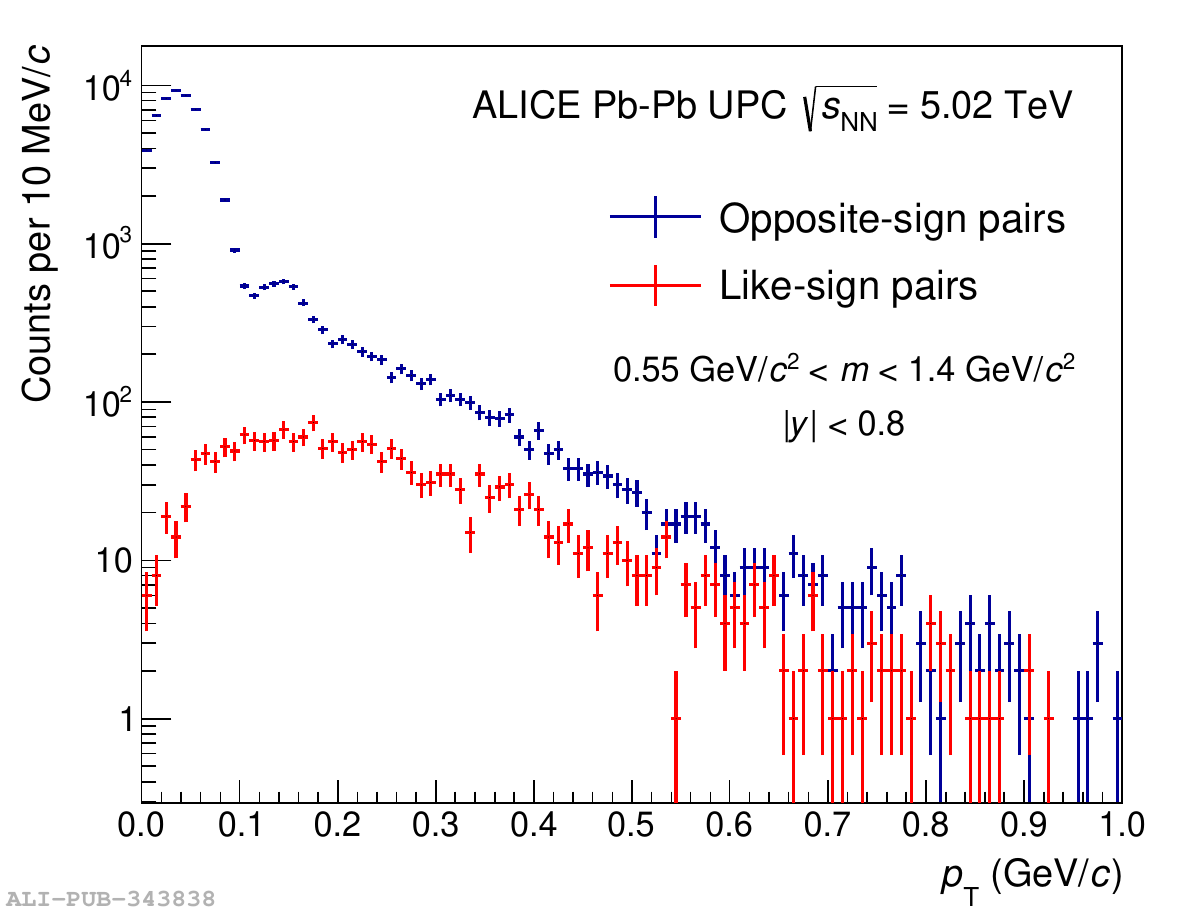}
		}
		\subfigure[]{
			\label{fig:rho-run3}
			\includegraphics[width=0.43\textwidth]{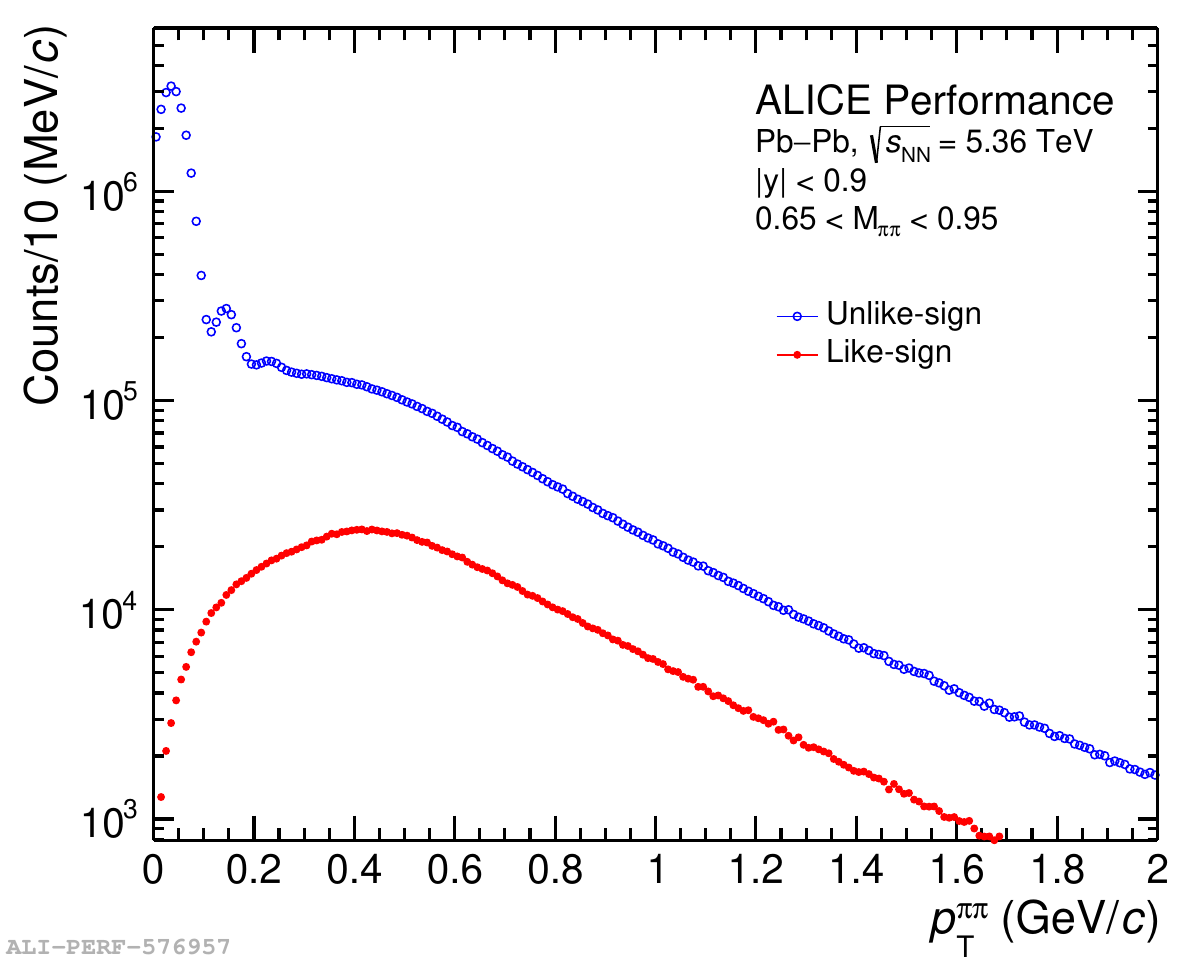}
		}\\ 
	\end{center}
	\caption{Transverse momentum distribution of $\pi\pi$ pairs selected in Run 2 (left panel) \cite{ALICE:2020ugp} and Run 3 (right panel), for UPC measurements in ALICE.}
	\label{fig:sl}
\end{figure}

\begin{figure}[thb!!]
\begin{center}
\includegraphics[width=0.55\columnwidth]{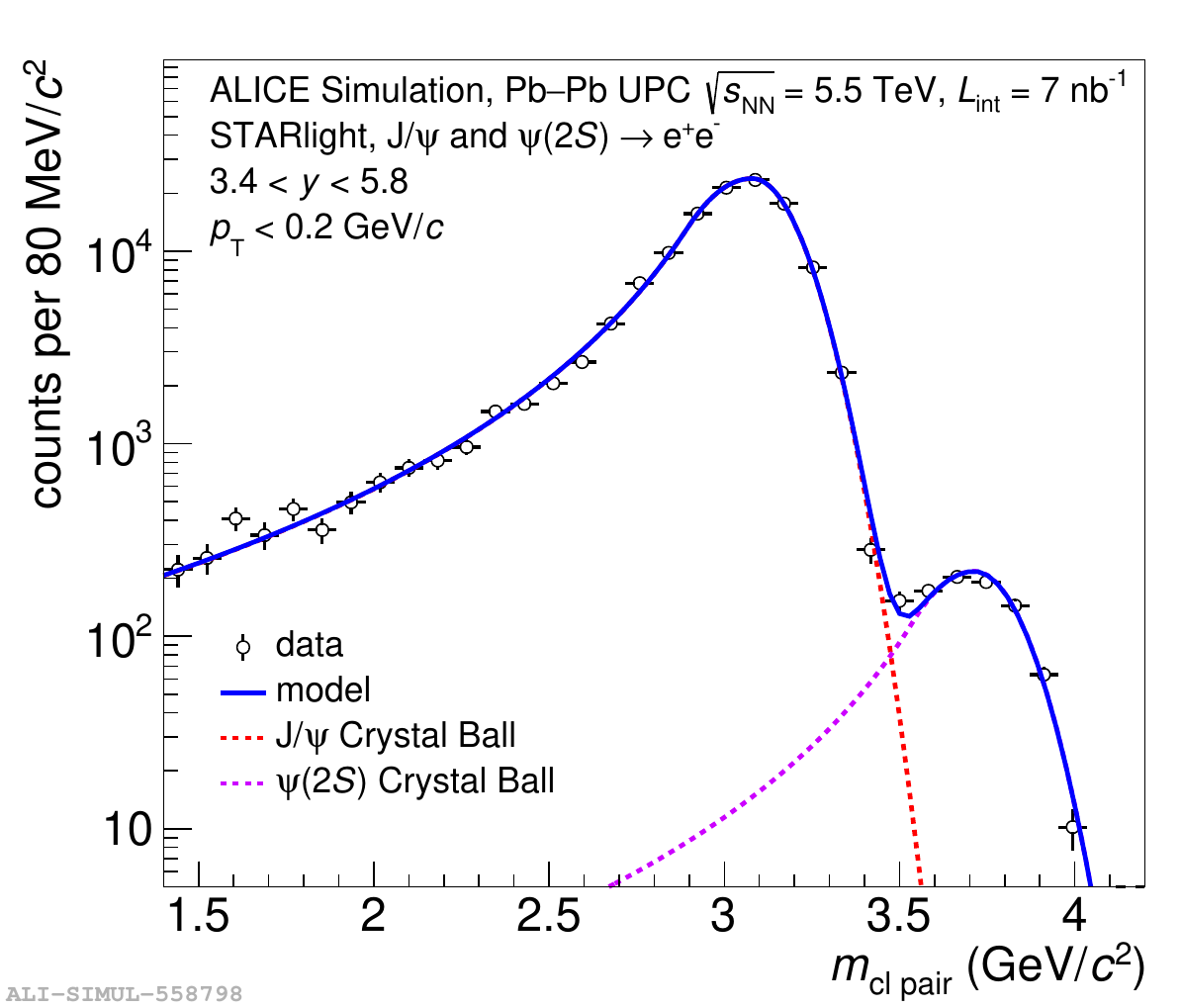}
\end{center}
\caption{\label{fig:focal} ALICE simulations using STARlight \cite{starlight} showing the potential of FoCal to measure photoproduced charmonia decaying into dielectrons.}
\end{figure}

In Run 3 (2022-2026) and Run 4 (starting in 2030) ALICE will collect significantly higher amounts of data \cite{Citron:2018lsq}, also in previously inaccessible rapidity regions, owing to the installation of new detectors, i.e. Muon Forward Tracker (MFT) in Run 3  and Forward Calorimeter (FoCal) in Run 4. The effects of the introduction of a continuous readout, are particularly evident using UPC selections.
While in Run 2 the sample contained about fifty thousand $\pi^+\pi^-$ candidates in the invariant-mass region of the \rhozero, as shown in Fig.~\ref{fig:rho-run2}, which were used in \cite{ALICE:2020ugp} and \cite{ALICE:2024ife}, the data set collected in Run 3 is already order of magnitudes larger in a similar invariant-mass region with UPC selections, as shown in Fig.~\ref{fig:rho-run3}. More precise and even more (multi-)differential measurements are expected with the new data sample.
In addition, the increased acceptance brought by the addition of MFT and FoCal will allow access to observables that are expected to significantly contribute to observation of the onset of the gluon saturation regime, as described in \cite{Bylinkin:2022wkm}. FoCal, which will be installed during Long Shutdown 3, will provide sensitivity to charmonia through their decay to dielectrons, as shown in Fig.~\ref{fig:focal}. This figure is an ALICE simulation produced using events from STARlight \cite{starlight}, where the \jpsi and \psip peaks are  clearly visible.
Finally, the ALICE detector in Run 3 and 4 should be able to collect enough statistics to measure e.g. light-by-light scattering \cite{Burmasov:2023yna, Burmasov:2021phy}, $\gamma\gamma\rightarrow\gamma\gamma$, and the anomalous magnetic moment in the tau sector \cite{Buhler:2022knp}, $\gamma\gamma\rightarrow\tau\tau$, allowing for measurements beyond the Standard Model with implications for axion-like particles (ALPs) \cite{Burmasov:2023yna} and SUSY \cite{Buhler:2022knp}. New opportunities will also arise through the usage of machine learning in UPCs, such as anomaly detection through autoencoders for the detection of exotic hadrons such as tetraquarks \cite{Goncalves:2021ytq} and pentaquarks, as described in \cite{Ragoni:2024ovv}.

\section{Conclusions}
The UPC program in ALICE brought about interesting new results, which cover a large range of phenomena, such as nuclear shadowing, gluon saturation and gluonic hotspots. The most recent measurements were however limited by their statistical uncertainty. The new incoming Run 3 and 4 data sets will provide abundant amounts of high quality data, allowing for more differential measurements. These large data sets will also permit the measurement of the $\gamma\gamma\rightarrow\gamma\gamma$ and $\gamma\gamma\rightarrow\tau\tau$ processes in ALICE, giving prospects for beyond the Standard Model investigations. The new detectors, i.e. MFT and FoCal in Run 3 and 4, respectively, will in addition allow access to new and exciting rapidity regions.


\bibliographystyle{JHEP}
\bibliography{biblio.bib}

\providecommand{\href}[2]{#2}\begingroup\raggedright\begin{thebibliography}{10}

\bibitem{Baltz:2007kq}
A.J.~Baltz, \emph{{The Physics of Ultraperipheral Collisions at the LHC}}, \href{https://doi.org/10.1016/j.physrep.2007.12.001}{\emph{Phys. Rept.} {\bfseries 458} (2008) 1} [\href{https://arxiv.org/abs/0706.3356}{{\ttfamily 0706.3356}}].

\bibitem{ALICE:2023jgu}
{\scshape ALICE} collaboration, \emph{{Energy dependence of coherent photonuclear production of J/\ensuremath{\psi} mesons in ultra-peripheral Pb-Pb collisions at $ \sqrt{{\textrm{s}}_{\textrm{NN}}} $ = 5.02 TeV}}, \href{https://doi.org/10.1007/JHEP10(2023)119}{\emph{JHEP} {\bfseries 10} (2023) 119} [\href{https://arxiv.org/abs/2305.19060}{{\ttfamily 2305.19060}}].

\bibitem{ALICE:2023mfc}
{\scshape ALICE} collaboration, \emph{{Exclusive and dissociative J/\ensuremath{\psi} photoproduction, and exclusive dimuon production, in p-Pb collisions at sNN=8.16\,\,TeV}}, \href{https://doi.org/10.1103/PhysRevD.108.112004}{\emph{Phys. Rev. D} {\bfseries 108} (2023) 112004} [\href{https://arxiv.org/abs/2304.12403}{{\ttfamily 2304.12403}}].

\bibitem{ALICE:2020ugp}
{\scshape ALICE} collaboration, \emph{{Coherent photoproduction of \rhozero vector mesons in ultra-peripheral \PbPb collisions at \fivenn}}, \href{https://doi.org/10.1007/JHEP06(2020)035}{\emph{JHEP} {\bfseries 06} (2020) 035} [\href{https://arxiv.org/abs/2002.10897}{{\ttfamily 2002.10897}}].

\bibitem{ALICE:2023kgv}
{\scshape ALICE} collaboration, \emph{{Photoproduction of K+K- Pairs in Ultraperipheral Collisions}}, \href{https://doi.org/10.1103/PhysRevLett.132.222303}{\emph{Phys. Rev. Lett.} {\bfseries 132} (2024) 222303} [\href{https://arxiv.org/abs/2311.11792}{{\ttfamily 2311.11792}}].

\bibitem{ALICE:2023gcs}
{\scshape ALICE} collaboration, \emph{{First measurement of the $|t|$ dependence of incoherent \jpsi photonuclear production}}, \href{https://doi.org/10.1103/PhysRevLett.132.162302}{\emph{Phys. Rev. Lett.} {\bfseries 132} (2024) 162302} [\href{https://arxiv.org/abs/2305.06169}{{\ttfamily 2305.06169}}].

\bibitem{Cepila:2023dxn}
J.~Cepila, J.G.~Contreras, M.~Matas and A.~Ridzikova, \emph{{Incoherent \jpsi production at large $|t|$ identifies the onset of saturation at the LHC}}, \href{https://doi.org/10.1016/j.physletb.2024.138613}{\emph{Phys. Lett. B} {\bfseries 852} (2024) 138613} [\href{https://arxiv.org/abs/2312.11320}{{\ttfamily 2312.11320}}].

\bibitem{ALICE:2024kjy}
{\scshape ALICE} collaboration, \emph{{Exclusive four pion photoproduction in ultraperipheral Pb-Pb collisions at $\sqrt{s_{\rm NN}} = 5.02$ TeV}},  \href{https://arxiv.org/abs/2404.07542}{{\ttfamily 2404.07542}}.

\bibitem{Klusek-Gawenda:2020gwa}
M.~Klusek-Gawenda and J.D.~Tapia~Takaki, \emph{{Exclusive Four-pion Photoproduction in Ultra-peripheral Heavy-ion Collisions at RHIC and LHC Energies}}, \href{https://doi.org/10.5506/APhysPolB.51.1393}{\emph{Acta Phys. Polon. B} {\bfseries 51} (2020) 1393} [\href{https://arxiv.org/abs/2005.13624}{{\ttfamily 2005.13624}}].

\bibitem{Reidt:2021tvq}
{\scshape ALICE} collaboration, \emph{{Upgrade of the ALICE ITS detector}}, \href{https://doi.org/10.1016/j.nima.2022.166632}{\emph{Nucl. Instrum. Meth. A} {\bfseries 1032} (2022) 166632} [\href{https://arxiv.org/abs/2111.08301}{{\ttfamily 2111.08301}}].

\bibitem{ALICE:2024ife}
{\scshape ALICE} collaboration, \emph{{Measurement of the impact-parameter dependent azimuthal anisotropy in coherent \rhozero photoproduction in \PbPb collisions at \fivenn}}, \href{https://doi.org/10.1016/j.physletb.2024.139017}{\emph{Phys. Lett. B} {\bfseries 858} (2024) 139017} [\href{https://arxiv.org/abs/2405.14525}{{\ttfamily 2405.14525}}].

\bibitem{Guzey:2013jaa}
V.~Guzey, M.~Strikman and M.~Zhalov, \emph{{Disentangling coherent and incoherent quasielastic \jpsi photoproduction on nuclei by neutron tagging in ultraperipheral ion collisions at the LHC}}, \href{https://doi.org/10.1140/epjc/s10052-014-2942-z}{\emph{Eur. Phys. J. C} {\bfseries 74} (2014) 2942} [\href{https://arxiv.org/abs/1312.6486}{{\ttfamily 1312.6486}}].

\bibitem{Baltz:2002pp}
A.J.~Baltz, S.R.~Klein and J.~Nystrand, \emph{{Coherent vector meson photoproduction with nuclear breakup in relativistic heavy ion collisions}}, \href{https://doi.org/10.1103/PhysRevLett.89.012301}{\emph{Phys. Rev. Lett.} {\bfseries 89} (2002) 012301} [\href{https://arxiv.org/abs/nucl-th/0205031}{{\ttfamily nucl-th/0205031}}].

\bibitem{Citron:2018lsq}
Z.~Citron et~al., \emph{{Report from Working Group 5}: {Future physics opportunities for high-density QCD at the LHC with heavy-ion and proton beams}}, \href{https://doi.org/10.23731/CYRM-2019-007.1159}{\emph{CERN Yellow Rep. Monogr.} {\bfseries 7} (2019) 1159} [\href{https://arxiv.org/abs/1812.06772}{{\ttfamily 1812.06772}}].

\bibitem{Bylinkin:2022wkm}
A.~Bylinkin, J.~Nystrand and D.~Tapia~Takaki, \emph{{Vector meson photoproduction in UPCs with FoCal}}, \href{https://doi.org/10.1088/1361-6471/acc419}{\emph{J. Phys. G} {\bfseries 50} (2023) 5} [\href{https://arxiv.org/abs/2211.16107}{{\ttfamily 2211.16107}}].

\bibitem{starlight}
S.~Klein, J.~Nystrand, J.~Seger, Y.~Gorbunov and J.~Butterworth, \emph{\text{STARlight:} \text{A Monte Carlo} simulation program for ultra-peripheral collisions of relativistic ions}, \href{https://doi.org/10.1016/j.cpc.2016.10.016}{\emph{Computer Physics Communications} {\bfseries 212} (2017) 258}.

\bibitem{Burmasov:2023yna}
N.A.~Burmasov, \emph{{Prospects of light-by-light scattering measurements and axion-like particle searches at the LHC}}, \href{https://doi.org/10.18721/JPM.161.247}{\emph{St. Petersburg Polytech. Univ. J. Phys. Math.} {\bfseries 16} (2023) 308}.

\bibitem{Burmasov:2021phy}
N.~Burmasov, E.~Kryshen, P.~Buehler and R.~Lavicka, \emph{{Upcgen: A Monte Carlo simulation program for dilepton pair production in ultra-peripheral collisions of heavy ions}}, \href{https://doi.org/10.1016/j.cpc.2022.108388}{\emph{Comput. Phys. Commun.} {\bfseries 277} (2022) 108388} [\href{https://arxiv.org/abs/2111.11383}{{\ttfamily 2111.11383}}].

\bibitem{Buhler:2022knp}
P.~B\"uhler, N.~Burmasov, R.~Lavi\v{c}ka and E.~Kryshen, \emph{{Feasibility study of tau-lepton anomalous magnetic moment measurements with ultra-peripheral collisions at the LHC}}, \href{https://doi.org/10.1051/epjconf/202226201021}{\emph{EPJ Web Conf.} {\bfseries 262} (2022) 01021}.

\bibitem{Goncalves:2021ytq}
V.P.~Gon\c{c}alves and B.D.~Moreira, \emph{{Fully - heavy tetraquark production by $\gamma\gamma$ interactions in hadronic collisions at the LHC}}, \href{https://doi.org/10.1016/j.physletb.2021.136249}{\emph{Phys. Lett. B} {\bfseries 816} (2021) 136249} [\href{https://arxiv.org/abs/2101.03798}{{\ttfamily 2101.03798}}].

\bibitem{Ragoni:2024ovv}
S.~Ragoni, J.~Seger and C.~Anson, \emph{{Zero-bias new particle searches using autoencoders in UPCs and diffractive events}},  \href{https://arxiv.org/abs/2411.00903}{{\ttfamily 2411.00903}}.

\end{thebibliography}\endgroup






\end{document}